\documentclass[aps,english,showpacs,twocolumn]{revtex4}
\usepackage{amssymb}
\usepackage{amsmath}
\usepackage{graphicx}
\usepackage{epsfig}

\begin{document}

\title{Fast transfer and efficient coherent separation of bound cluster in
the extended Hubbard model}
\author{L. Jin, and Z. Song}
\email{songtc@nankai.edu.cn}
\affiliation{School of Physics, Nankai University, Tianjin 300071, China}

\begin{abstract}
We study the formation and dynamics of the bound pair (BP) and bound triple
(BT) in strongly correlated extended Hubbard model for both Bose and Fermi
systems. We find that the bandwidths of the BP and BT gain significantly
when the on-site and nearest-neighbor interaction strengths reach the
corresponding resonant points. This allows the fast transfer and efficient
coherent separation of the BP and BT. The exact result shows that the
success probability of the coherent separation is unity in the optimal
system. In Fermi system, this finding can be applied to create distant
entanglement without the need of temporal control and measurement process.
\end{abstract}

\maketitle

%03.65.Ge Solutions of wave equations: bound states
%05.30.Jp Boson systems
%03.65.Nk Scattering theory
%03.67.-a Quantum information

\section{Introduction}

Ultra-cold atoms have turned out to be an ideal playground for testing
few-particle fundamental physics and for their increasing technological
applications. The experimental observation of atomic bound pair (BP) in
optical lattice \cite{Winkler} stimulates many experimental and theoretical
investigations in strongly correlated boson systems \cite{Mahajan,
Petrosyan, Creffield, Kuklov, Folling, Zollner, ChenS, Valiente, JLBP,
MVExtendB, MVTB}. The essential physics of the proposed BP is that, the
periodic potential suppresses the single particle tunneling across the
barrier, a process that would lead to a decay of the pair. Such kind of BP
therefore cannot get migration speed through the lattice. In addition, the
dynamics of the pair separation and formation is also of interest in both
fundamental and application aspects.

In this article we study the influence of the nearest-neighbor (NN)
interaction on the few-boson bound state. We will show that, the resonant NN
interaction can also induce the BP and bound triple (BT) states. These bound
states are distinct from the known ones, as their bandwidths are comparable
to that of a single particle. This feature allows the full overlap between
the scattering band of a single particle and the band of a BP or a BT,
leading to the coherent separation and combination processes. We propose
relatively simple schemes to implement such processes. The exact result
shows that the success probability of the coherent separation is unity for a
BP while approaches to unity for a BT in the optimal systems. Applying the
scheme to the extended Fermi-Hubbard system, the coherent separation process
can be utilized to create long-distance entanglement without the need of
temporal control and measurement process.

\section{Fast bound pair dynamics}

Let us start by analyzing in detail the two-particle problem in an extended
Bose-Hubbard model, which is simpler than the Fermi one but shares the
similar features for the issue concerned. The Hamiltonian is written as
follows:
\begin{equation}
H_{\text{B}}=-t\overset{N}{\sum_{i=1}}\left( a_{i}^{\dag }a_{i+1}+\text{H.c.}%
\right) +\frac{U}{2}\overset{N}{\sum_{i=1}}n_{i}(n_{i}-1)+V\overset{N}{%
\sum_{i=1}}n_{i}n_{i+1},  \label{H}
\end{equation}%
where $a_{i}^{\dag }$ is the creation operator of the boson at the $i$th
site, the tunneling strength, on-site and NN interactions between bosons are
denoted by $t$, $U$ and $V$. For the sake of clarity and simplicity, we only
consider odd-site system with $N=2N_{0}+1$, and periodic boundary conditions
$a_{N+1}=a_{1}$. In this article, we only concern the one-dimensional
system. Nevertheless the conclusion can be extended to high-dimensional
system.

First of all, a state in the two-particle Hilbert space, as shown in Ref.
\cite{JLBP}, can be written as $\left\vert \psi _{k}\right\rangle =$ $%
\sum_{k,r}f^{k}(r)\left\vert \phi _{r}^{k}\right\rangle ,$ with\
\begin{eqnarray}
\left\vert \phi _{0}^{k}\right\rangle &=&\frac{1}{\sqrt{2N}}e^{i\frac{k}{2}%
}\sum_{j}e^{ikj}\left( a_{j}^{\dag }\right) ^{2}\left\vert \text{vac}%
\right\rangle , \\
\left\vert \phi _{r}^{k}\right\rangle &=&\frac{1}{\sqrt{N}}e^{i\frac{k(r+1)}{%
2}}\sum_{j}e^{ikj}a_{j}^{\dag }a_{j+r}^{\dag }\left\vert \text{vac}%
\right\rangle ,
\end{eqnarray}%
here $\left\vert \text{vac}\right\rangle $\ is the vacuum state for the
boson operator $a_{i}$. $k=2\pi n/N$, $n\in \lbrack 1,N]$ denotes the
momentum, and $r\in \lbrack 1,N_{0}]$ is the distance between the two
particles. Due to the translational symmetry of the present system, the Schr%
\"{o}dinger equations for $f^{k}(r)$, $r\in \lbrack 0,N_{0}]$ are easily
shown to be
\begin{equation}
\begin{split}
& T_{r}^{k}f^{k}(r+1)+T_{r-1}^{k}f^{k}(r-1) \\
& +[U\delta _{r,0}+V\delta _{r,1}+\left( -1\right) ^{n}T_{r}^{k}\delta
_{r,N_{0}}-\varepsilon _{k}]f^{k}(r)=0,
\end{split}
\label{Schrodinger}
\end{equation}%
where $T_{r}^{k}=-2\sqrt{2}t\cos (k/2)$ for $r=0$, and $-2t\cos (k/2)$ for $%
r\neq 0$, respectively. Besides, we also have the boundary conditions $%
f^{k}(-1)=$ $f^{k}(N_{0}+1)=0$. Note that for an arbitrary $k$, the solution
of (\ref{Schrodinger}) is equivalent to that of the single-particle $N_{0}+1$%
-site tight-binding chain system with NN hopping amplitude $T_{j}^{k}$,\
on-site potentials $U$, $V$ and $-2t\cos (k/2)$\ at $0$th, $1$th and $N_{0}$%
th sites respectively. Obviously, in each $k$-invariant subspace, there are
three types of bound states arising from the on-site potentials under the
following conditions. In the case with $\left\vert U-V\right\vert \gg t$,\
the particle can be localized at either $0$th\ or $1$th\ site, corresponding
to (i) the on-site BP state, or (ii) the NN BP state. Interestingly, in the
case of resonance $U=V$, and $\left\vert U\right\vert $, $\left\vert
V\right\vert \gg t$, the particle can be in the bonding state (or
anti-bonding state) between $0$th\ and $1$th\ sites, corresponding to a new
BP state, called (iii) the resonant BP (RBP) state. In this article,
hereafter we forcus on this type of bound state and refer RBP as BP. All the
$N$ bound states of (iii), indexed by $k$,\ constitute a bound-pair band.

In previous works \cite{JLBP, MVExtendB}, the bound states of (i) and (ii)
were well investigated and the corresponding bound-pair bandwidths are of $%
t^{2}/U$ and $t^{2}/V\ $order. Then they can be regarded as stationary
comparing to the single particle in the strongly correlated limit. In order to analyze the
dynamics of the BP of type (iii), we pursue the solution of Eq. (\ref%
{Schrodinger}) in the case with $U=V$, and $\left\vert U\right\vert $, $%
\left\vert V\right\vert \gg t$,\ via\ the Bathe-ansatz method. The BP states
have the form

\begin{equation}
f^{k}(r)\simeq \eta _{k}\left\{
\begin{array}{c}
-\sqrt{2}\epsilon _{k}/\left( \varepsilon _{k}-U\right) \text{,\ }(r=0), \\
\left[ -\text{sgn}\left( U/\epsilon _{k}\right) \right] ^{\left( r-1\right)
}\left( \xi _{k}\right) ^{-\left( r-1\right) }\text{,}%
\end{array}%
\right.  \label{f(r)}
\end{equation}%
with the spectrum

\begin{equation}
\varepsilon _{k}\simeq U\pm \sqrt{2}\epsilon _{k}+\frac{\left( \epsilon
_{k}\right) ^{2}}{2U},  \label{e_k}
\end{equation}%
where

\begin{eqnarray}
\eta _{k}^{-2} &=&\frac{\left\vert \varepsilon _{k}\right\vert }{2\sqrt{%
\varepsilon _{k}^{2}-4\left( \epsilon _{k}\right) ^{2}}}+\frac{2\left(
\epsilon _{k}\right) ^{2}}{\left( \varepsilon _{k}-U\right) ^{2}}+\frac{1}{2}%
, \\
\xi _{k} &=&\left( \left\vert \varepsilon _{k}\right\vert +\sqrt{\varepsilon
_{k}^{2}-4\left( \epsilon _{k}\right) ^{2}}\right) /\left( 2\left\vert
\epsilon _{k}\right\vert \right) ,
\end{eqnarray}%
and

\begin{equation}
\epsilon _{k}=-2t\cos \left( k/2\right) ,
\end{equation}%
is partial dispersion relation of a single particle. We note that neglecting
the terms with $t^{2}/U$, the two branches of the spectrum combine into a BP
band and the corresponding eigenfunction and energy can be rewritten as%
\begin{equation}
\left\vert \psi _{k}^{\text{BP}}\right\rangle =\sum_{j}\frac{e^{ik\left(
j+1/2\right) }}{\sqrt{2N}}\left[ \left( a_{j}^{\dag }\right) ^{2}/\sqrt{2}%
\mp e^{i\frac{k}{2}}a_{j}^{\dag }a_{j+1}^{\dag }\right] \left\vert \text{vac}%
\right\rangle ,  \label{PSI_RBP}
\end{equation}%
\begin{equation}
\varepsilon _{k}^{\text{BP}}=U\pm 2\sqrt{2}t\cos \left( k/2\right) .
\label{E_RBP}
\end{equation}%
Observing the above expressions, we find that it can be regarded as a plane
wave of the composite particle, BP. Actually, this can be easily understood
in terms of the equivalent effective Hamiltonian, which can be obtained in
the quasi-invariant subspace spanned by the basis $\{\underline{\left\vert
l\right\rangle },l\in \left[ 1,2N\right] \}$ with diagonal energy $U$. The
set of basis $\{\underline{\left\vert l\right\rangle }\}$ is defined as%
\begin{equation}
\underline{\left\vert l\right\rangle }\equiv \left\{
\begin{array}{c}
\left( a_{l/2}^{\dag }\right) ^{2}/\sqrt{2}\left\vert \text{vac}%
\right\rangle \text{, }(\text{even }l) \\
a_{\left( l-1\right) /2}^{\dag }a_{\left( l+1\right) /2}^{\dag }\left\vert
\text{vac}\right\rangle \text{,\ }(\text{odd }l)%
\end{array}%
\right. ,  \label{BP_basis}
\end{equation}%
The effective Hamiltonian of the ring, restricted to the basis $\{\underline{%
\left\vert l\right\rangle }\}$ and shifted by $U$, reads%
\begin{equation}
\mathcal{H}_{\text{BP}}=-\sqrt{2}t\overset{2N}{\sum_{l=1}}\left( \underline{%
\left\vert l\right\rangle }\underline{\left\langle l+1\right\vert }+\text{%
H.c.}\right) ,  \label{H_ring}
\end{equation}%
which describes a free Bloch particle in a single band. However, this BP
state is distinct from the previous on-site and NN BP because of its large
bandwidth, $4\sqrt{2}t$. It shows that the speed of a BP wavepacket must be
in the range of $0\sim 2\sqrt{2}t$, which is the reason why we call it fast
bound pair. This is very significant for quantum technologies in the
following two aspects: (i) The BP wavepacket can be a new candidate as
quantum information carrier due to its fast speed. (ii) The bandwidth
matching between BP and a single particle may lead to the coherent
separation of the BP.

%%%%%%%%%%%%%%%%%%%%%%%%%%%%%%%%%%%%%%%%%%%%%%%%%%%%%%%%%%%%%%%%%%%%%%%%
\begin{figure}[tbp]
\includegraphics[ bb=89 348 514 716, width=6.0 cm, clip]{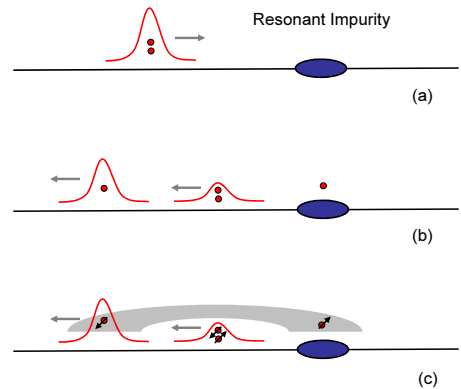}
\caption{(Color online) Schematic illustration of the coherent separation
process of a BP in the proposed optimal system. It consists of a uniform
chain and the embedded resonant impurity. (a) An incident BP wavepacket
comes from the left and scatters off the resonant impurity. (b) The
reflected waves contain the BP and the single particle wavepackets with
different speeds. A single particle resides at the impurity. For an optimal
incident wavepacket, the success probability of the coherent separation is
unity. (c) Appling the previous scheme to a singlet femionic BP, it is found
that the separated two fermions are maximally entangled.}
\label{BP}
\end{figure}

%%%%%%%%%%%%%%%%%%%%%%%%%%%%%%%%%%%%%%%%%%%%%%%%%%%%%%%%%%%%%%%%%%%%%%%%

\section{Coherent separation of bound pair}

Considering a system of (\ref{H}) with the additional chemical potential $%
\mu =U$,\ the scattering band of a single particle fully overlaps the band
of a BP (\ref{e_k}). This fact indicates that a BP may break in the
scattering process by a resonant impurity. In order to demonstrate a perfect
process of the BP separation, we propose an\ optimal system which is a
uniform chain with a resonant impurity embedded in. The uniform chain acts
as a channel for the transport of a BP wave, under the conditions $U=V$, and
$\left\vert U\right\vert $, $\left\vert V\right\vert \gg t$.\textbf{\ }The
resonant impurity consists of two sites with the tunneling strength\textbf{\
}$t_{0}$, one site of which has far-off resonant on-site interaction $U_{s}$
satisfying $\left\vert U_{s}-U\right\vert \gg t$ and another site has
resonant chemical potential $\mu =U$.

This setup is described by the Hamiltonian

\begin{eqnarray}
&&H_{\text{BPS}}=-\left( t\overset{\infty }{\sum_{i=-\infty }}a_{i}^{\dag
}a_{i+1}+\left( t_{0}-t\right) a_{0}^{\dag }a_{1}\right) +\text{H.c.}
\label{H_s} \\
&&+U\overset{\infty }{\sum_{i=-\infty }}n_{i}(\frac{n_{i}}{2}+n_{i+1}-1)+%
\frac{U_{s}-U}{2}n_{0}(n_{0}-1)+Un_{1}.  \notag
\end{eqnarray}%
To investigate the scattering process of an input BP wavepacket from the
left, one can establish an equivalent effective Hamiltonian in the
quasi-invariant subspace spanned by the following basis $\{\underline{%
\left\vert l\right\rangle },l\in \left( -\infty ,\infty \right) \}$ defined
as%
\begin{equation}
\underline{\left\vert l\right\rangle }\equiv \left\{
\begin{array}{c}
\text{Eqs. (\ref{BP_basis}), }(l<0) \\
a_{-l-1}^{\dag }a_{1}^{\dag }\left\vert \text{vac}\right\rangle \text{,\ }%
(l\geqslant 0)%
\end{array}%
\right. .  \label{basis_s}
\end{equation}%
Acting $H_{\text{BPS}}$\ on the subspace, the equivalent effective
Hamiltonian reads%
\begin{equation}
\mathcal{H}_{\text{BPS}}=-t\left( \sqrt{2}\overset{-1}{\sum_{l=-\infty }}%
\overset{\infty }{+\sum_{l=1}}\right) \underline{\left\vert l-1\right\rangle
}\underline{\left\langle l\right\vert }-t_{0}\underline{\left\vert
-1\right\rangle }\underline{\left\langle 0\right\vert }+\text{H.c.}
\label{H_eff}
\end{equation}%
which describes a two connected semi-infinite chains with different hopping
constants. The on-site interaction $U_{s}$ ensures $H_{\text{BPS}}$ to be a
linear chain, which would be of benefit to enhance the success probability
of the coherent separation.

The separation process of a BP wavepacket is illustrated in Fig. \ref{BP}.
After scattering, a part of the BP wavepacket is reflected by the impurity,
while the other part of it is separated into two independent particles: One
of the particle resides at the site with chemical potential $U$, while the
other particle is reflected to the left with higher speed than the reflected
BP wavepacket. In the aid of the effective Hamiltonian (\ref{H_eff}), the
previous process can be reduced to a simple single particle scattering
problem: An incident wavepacket is scattered by the joint of the two
semi-infinite chains. The reflecting wave represents the reflecting BP wave,
while the transmitting wave represents the separated reflecting particle. In
this sense, the coherent separation probability is equal to the transmission
coefficient of the effective Hamiltonian, which can be obtained exactly via
Green' function or Bethe-ansatz method \cite{Datta,YangAs,JLTrans}.

\begin{figure}[tbp]
\includegraphics[ bb=22 304 577 765, width=6.0 cm, clip]{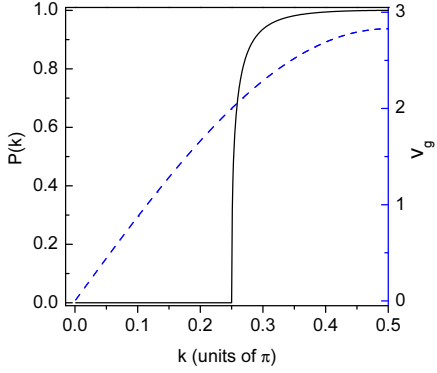}
\caption{(Color online) Plots of the separation probability $P\left(
k\right) $ (solid line) and the group velocity $v_{g}$ (dashed line) as
functions of the momentum of the incident wavepacket. It shows that both of
them get their maxima at $\protect\pi /2$.}
\label{Pk}
\end{figure}

For an incident plane wave of $k_{0}$, the coherent separation probability $%
P\left( k_{0}\right) $ for a chosen system of $t_{0}=\sqrt[4]{2}t$ is

\begin{equation}
P\left( k_{0}\right) =2\left[ 1+\frac{1-\sqrt{2}\cos ^{2}\left( k_{0}\right)
}{\sin k_{0}\sqrt{\left\vert \cos \left( 2k_{0}\right) \right\vert }}\right]
^{-1},  \label{T}
\end{equation}%
for $k_{0}\in (\pi /4,\pi /2]$\ and $P\left( k_{0}\right) =0$ for $k_{0}\in
(0,\pi /4]$. The sudden death of $P\left( k_{0}\right) $ within the region $%
(0,\pi /4]$ is due to the mismatch between energies of the two sides of the
joint. On the other hand,$\ P\left( k_{0}\right) $ also represents the
particle resident population at the impurity. The profile of $P\left(
k_{0}\right) $ is plotted in Fig. \ref{Pk}. It indicates that $P\left(
k_{0}\right) $ can reach $1.0$ at $k_{0}=\pi /2$.

In practice, this process can be implemented via an incident wavepacket
instead of a plane wave. Actually, the typical wavepacket, a Gaussian wave
packet in the space $\{\underline{\left\vert l\right\rangle }\}$ can be
constructed as%
\begin{equation}
\left\vert \Phi (k_{0},N_{c})\right\rangle =\frac{1}{\sqrt{\Omega }}%
\sum_{l}e^{-\frac{^{\alpha ^{2}}}{2}(l-N_{c})^{2}+ik_{0}l}\underline{%
\left\vert l\right\rangle },  \label{Gaussian WP}
\end{equation}%
where $\Omega $ is the normalization factor. Here $N_{c}$ is the center of
it in the space $\{\underline{\left\vert l\right\rangle }\}$, $k_{0}$ is the
momentum of it. The corresponding group velocity in the space $\{\underline{%
\left\vert l\right\rangle }\}$\ is $v_{g}=-2\sqrt{2}t\sin \left(
k_{0}\right) $ which is also plotted in Fig. \ref{Pk}. It shows that for the
fastest wavepacket with momentum $\pm \pi /2$, the success probability of
the coherent separation can approach to unity. In addition, it is shown that
such a wavepacket is the most robust against spreading \cite{KimImpurity}.
We define it as the optimal BP wavepacket to demonstrate the perfect
coherent separation. In the original system, it\ can be written as

\begin{eqnarray}
\left\vert \phi (\pm ,h_{c})\right\rangle &\simeq &\frac{1}{\sqrt{\Omega }}%
\sum_{j}\left( -1\right) ^{j}e^{-2\alpha ^{2}\left( j-h_{c}\right) ^{2}}
\notag \\
&&\times \left[ \left( a_{j}^{\dag }\right) ^{2}/\sqrt{2}\pm ia_{j}^{\dag
}a_{j+1}^{\dag }\right] \left\vert \text{vac}\right\rangle ,  \label{BP_PI}
\end{eqnarray}%
which is obtained from (\ref{Gaussian WP}) with wide width $\left( \alpha
\ll 1\right) $ and $k_{0}=\pm \pi /2$\ by the mapping rule (\ref{basis_s}).
Here symbols $\pm $ and $h_{c}$\ denote the moving direction (sgn$\left( \pm
\pi /2\right) $) and the center of the wavepacket.\ The corresponding single
particle wave packet has the form%
\begin{equation}
\left\vert \varphi (\pm ,h_{c})\right\rangle =\frac{1}{\sqrt{\Omega }}%
\sum_{j}e^{-\frac{^{\alpha ^{2}}}{2}(j-h_{c})^{2}\pm i\frac{\pi }{2}%
j}a_{j}^{\dag }\left\vert \text{vac}\right\rangle .  \label{single_P}
\end{equation}%
Note that $\left\vert \phi (\pm ,h_{c})\right\rangle $ is narrower and
slower than that of $\left\vert \varphi (\pm ,h_{c})\right\rangle $. This is
also illustrated in Fig. \ref{BP} (b) and (c).

Thus, in the optimal case, the perfect scattering process can be expressed as

\begin{eqnarray}
&&\left\vert \phi (+,-\infty )\right\rangle \left\vert \text{vac}%
\right\rangle \Longrightarrow  \notag \\
&&r\left\vert \phi (-,-\infty )\right\rangle \left\vert \text{vac}%
\right\rangle +t\left\vert \varphi (-,-\infty )\right\rangle a_{1}^{\dag
}\left\vert \text{vac}\right\rangle ,  \label{sep}
\end{eqnarray}%
and the corresponding inverse process as

\begin{eqnarray}
&&\left\vert \varphi (+,-\infty )\right\rangle a_{1}^{\dag }\left\vert \text{%
vac}\right\rangle \Longrightarrow  \notag \\
&&r\left\vert \varphi (-,-\infty )\right\rangle a_{1}^{\dag }\left\vert
\text{vac}\right\rangle +t\left\vert \phi (-,-\infty )\right\rangle
\left\vert \text{vac}\right\rangle ,  \label{comb}
\end{eqnarray}%
where $\Longrightarrow $\ represents the time evolution. Eqs. (\ref{sep})
and (\ref{comb})\textbf{\ }represent the reversibility of the processes,
coherent separation and combination, due to the time-reversal symmetry of
the system. Here $r$ and $t$ represent the reflection and transmission
amplitudes. Then the success probability of the coherent separation is $%
\left\vert t\right\vert ^{2}$. Numerical simulation is performed to evaluate
the separation probability. For an optimal Gaussian wavepacket with $\alpha
=0.01$, we have $\left\vert t\right\vert ^{2}=1.00$. Hence we conclude that
the coherent separation scheme is indeed very efficient due to its three
advantages, fast transfer, robustness and high efficiency.

%%%%%%%%%%%%%%%%%%%%%%%%%%%%%%%%%%%%%%%%%%%%%%%%%%%%%%%%%%%%%%%%%%%%%%%%

%%%%%%%%%%%%%%%%%%%%%%%%%%%%%%%%%%%%%%%%%%%%%%%%%%%%%%%%%%%%%%%%%%%%%%%%

\section{Creation of distant entanglement}

Now we turn to analyze the similar problem in the extended Fermi-Hubbard
model. The Hamiltonian reads%
\begin{equation}
\begin{split}
& H_{\text{F}}=-\left( \overset{\infty }{t\sum_{i=-\infty ,\sigma }}%
c_{i\sigma }^{\dag }c_{i+1\sigma }+\left( t_{0}-t\right) c_{0\sigma }^{\dag
}c_{1\sigma }\right) +\text{H.c.} \\
& +U\overset{\infty }{\sum_{i=-\infty }}(n_{i\uparrow }n_{i\downarrow
}+n_{i}n_{i+1})+\left( U_{s}-U\right) n_{0\uparrow }n_{0\downarrow }+Un_{1},
\end{split}
\label{H_F}
\end{equation}%
where $c_{i\sigma }^{\dagger }$ is the creation operator of fermion at site $%
i$ with spin $\sigma =\uparrow ,\downarrow $ and $n_{i}=n_{i\uparrow
}+n_{i\downarrow }$. For similar two-particle problem, all the previous
conclusions for the Bose system are valid for the Fermi system under the
following mapping rule:

\begin{equation}
\begin{split}
& \left( a_{j}^{\dag }\right) ^{2}/\sqrt{2}\left\vert \text{vac}%
\right\rangle \rightarrow c_{j\uparrow }^{\dag }c_{j\downarrow }^{\dag
}\left\vert \text{vac}\right\rangle , \\
& a_{i}^{\dag }a_{j}^{\dag }\left\vert \text{vac}\right\rangle \rightarrow
\left( c_{i\uparrow }^{\dag }c_{j\downarrow }^{\dag }-c_{i\downarrow }^{\dag
}c_{j\uparrow }^{\dag }\right) /\sqrt{2}\left\vert \text{vac}\right\rangle .
\end{split}%
\end{equation}%
Accordingly we represent the wavepacket as

\begin{eqnarray}
\left\vert \phi (\pm ,N_{c})\right\rangle &\rightarrow &\left\vert \phi
_{\uparrow \downarrow }(\pm ,N_{c})\right\rangle , \\
\left\vert \varphi (\pm ,N_{c})\right\rangle &\rightarrow &\left\vert
\varphi _{\sigma }(\pm ,N_{c})\right\rangle .  \notag
\end{eqnarray}%
Nevertheless, from the process

\begin{eqnarray}
&&\left\vert \phi _{\uparrow \downarrow }(+,-\infty )\right\rangle
\left\vert \text{vac}\right\rangle \Longrightarrow r\left\vert \phi
_{\uparrow \downarrow }(-,-\infty )\right\rangle \left\vert \text{vac}%
\right\rangle  \label{singlet} \\
&&+\frac{t}{\sqrt{2}}\left( \left\vert \varphi _{\uparrow }(-,-\infty
)\right\rangle c_{1\downarrow }^{\dag }-\left\vert \varphi _{\downarrow
}(-,-\infty )\right\rangle c_{1\uparrow }^{\dag }\right) \left\vert \text{vac%
}\right\rangle  \notag
\end{eqnarray}%
and the corresponding inverse process

\begin{eqnarray}
&&\left( \left\vert \varphi _{\uparrow }(+,-\infty )\right\rangle
c_{1\downarrow }^{\dag }-\left\vert \varphi _{\downarrow }(+,-\infty
)\right\rangle c_{1\uparrow }^{\dag }\right) \left\vert \text{vac}%
\right\rangle \Longrightarrow \\
&&+r\left( \left\vert \varphi _{\uparrow }(-,-\infty )\right\rangle
c_{1\downarrow }^{\dag }-\left\vert \varphi _{\downarrow }(-,-\infty
)\right\rangle c_{1\uparrow }^{\dag }\right) \left\vert \text{vac}%
\right\rangle  \notag \\
&&\sqrt{2}t\left\vert \phi _{\uparrow \downarrow }(-,-\infty )\right\rangle
\left\vert \text{vac}\right\rangle ,  \notag
\end{eqnarray}%
we can see that the two separated fermions have distinct feature from that
of bosons, which arises from the fact that, state $a_{i}^{\dag }a_{j}^{\dag
}\left\vert \text{vac}\right\rangle $ is a separable state while state $%
(c_{i\uparrow }^{\dag }c_{j\downarrow }^{\dag }-c_{i\downarrow }^{\dag
}c_{j\uparrow }^{\dag })/\sqrt{2}\left\vert \text{vac}\right\rangle $\ is
the maximally entangled state. Utilizing this scheme, one can obtain a pair
of almost maximally entangled nodes, shared by Alice and Bob. This can be
used for perfect quantum transport via teleportation.

In addition, when we consider the combination process of two fermions with
different spin orientations, it becomes a little complicated. We will start
our analysis from the simplest case in two-particle problem, two parallel
fermions. It is equivalent to the two spinless fermions system, in which BP
no longer exists. The corresponding process are%
\begin{equation}
\left\vert \varphi _{\uparrow }(+,-\infty )\right\rangle c_{1\uparrow
}^{\dag }\left\vert \text{vac}\right\rangle \Longrightarrow \left\vert
\varphi _{\uparrow }(-,-\infty )\right\rangle c_{1\uparrow }^{\dag
}\left\vert \text{vac}\right\rangle ,  \label{upup}
\end{equation}%
which represents the complete reflection of the incident wavepacket. Due to
the SU(2) symmetry of the Hamiltonian (\ref{H_s}), we have%
\begin{eqnarray}
&&\left\vert \varphi _{\uparrow }(+,-\infty )\right\rangle c_{1\downarrow
}^{\dag }\left\vert \text{vac}\right\rangle +\left\vert \varphi _{\downarrow
}(+,-\infty )\right\rangle c_{1\uparrow }^{\dag }\left\vert \text{vac}%
\right\rangle \Longrightarrow  \notag \\
&&\left\vert \varphi _{\uparrow }(-,-\infty )\right\rangle c_{1\downarrow
}^{\dag }\left\vert \text{vac}\right\rangle +\left\vert \varphi _{\downarrow
}(-,-\infty )\right\rangle c_{1\uparrow }^{\dag }\left\vert \text{vac}%
\right\rangle .  \label{updown}
\end{eqnarray}%
Eqs. (\ref{singlet}), (\ref{upup}) and (\ref{updown}) lead to%
\begin{equation}
\begin{split}
& \left[ \alpha \left\vert \varphi _{\uparrow }(+,-\infty )\right\rangle
+\beta \left\vert \varphi _{\downarrow }(+,-\infty )\right\rangle \right]
c_{1\uparrow }^{\dag }\left\vert \text{vac}\right\rangle \Longrightarrow \\
& \alpha \left\vert \varphi _{\uparrow }(-,-\infty )\right\rangle
c_{1\uparrow }^{\dag }\left\vert \text{vac}\right\rangle -\frac{\beta t}{%
\sqrt{2}}\left\vert \phi _{\uparrow \downarrow }(-,-\infty )\right\rangle
\left\vert \text{vac}\right\rangle \\
& +\frac{\beta }{2}\left( \left\vert \varphi _{\uparrow }(-,-\infty
)\right\rangle c_{1\downarrow }^{\dag }+\left\vert \varphi _{\downarrow
}(-,-\infty )\right\rangle c_{1\uparrow }^{\dag }\right) \left\vert \text{vac%
}\right\rangle \\
& -\frac{\beta r}{2}\left( \left\vert \varphi _{\uparrow }(-,-\infty
)\right\rangle c_{1\downarrow }^{\dag }-\left\vert \varphi _{\downarrow
}(-,-\infty )\right\rangle c_{1\uparrow }^{\dag }\right) \left\vert \text{vac%
}\right\rangle .
\end{split}
\label{arb}
\end{equation}%
The merit of the above scheme lies in the property of the natural time
evolution process in an always on system\ without the need of temporal
control and measurement process.

\section{Bound triple}

We now consider the three-particle bound state, BT, in the extended
Bose-Hubbard model. For the sake of simplicity, we directly investigate the
system in the limit $\left\vert V\right\vert \gg t$, $\left\vert
U\right\vert $, where the perturbation method is applicable and provides a
clear physical picture. The BT state is constructed by the three-particle
cluster in the configurations, $a_{i-1}^{\dag }a_{i}^{\dag }a_{i+1}^{\dag
}\left\vert \text{vac}\right\rangle $ and $(a_{i}^{\dag })^{2}a_{i\pm
1}^{\dag }\left\vert \text{vac}\right\rangle $, which possess the same
diagonal energy $2V$. Since the transition strength between them (or the
bandwidth of the new composite particle) is of the order of $t$, there may
exists fast transfer mode in such a system. Accordingly, its dynamics obeys
the following effective Hamiltonian

\begin{eqnarray}
\mathcal{H}_{\text{BT}} &=&-\sqrt{2}t\overset{\infty }{\sum_{i=-\infty }}%
\left( \underline{\left\vert 3i-1\right\rangle }\underline{\left\langle
3i\right\vert }+\underline{\left\vert 3i\right\rangle }\underline{%
\left\langle 3i+1\right\vert }\right.  \label{H_tri} \\
&&\left. +\sqrt{2}\underline{\left\vert 3i+1\right\rangle }\underline{%
\left\langle 3i+2\right\vert }+\text{H.c.}\right)  \notag
\end{eqnarray}%
which is obtained based on the quasi-invariant subspace spanned by the basis
$\{\underline{\left\vert l\right\rangle }\}$%
\begin{equation}
\underline{\left\vert l\right\rangle }\equiv \left\{
\begin{array}{l}
a_{l/3-1}^{\dag }a_{l/3}^{\dag }a_{l/3+1}^{\dag }\left\vert \text{vac}%
\right\rangle \text{, }(\text{mod}\left( l,3\right) =0), \\
\multicolumn{1}{c}{\left( a_{\left( l-1\right) /3}^{\dag }\right)
^{2}a_{\left( l+2\right) /3}^{\dag }/\sqrt{2}\left\vert \text{vac}%
\right\rangle \text{, }(\text{mod}\left( l-1,3\right) =0),} \\
\multicolumn{1}{c}{a_{\left( l-2\right) /2}^{\dag }\left( a_{\left(
l+1\right) /2}^{\dag }\right) ^{2}/\sqrt{2}\left\vert \text{vac}%
\right\rangle \text{, }(\text{mod}\left( l-2,3\right) =0).}%
\end{array}%
\right.  \label{Tri_basis}
\end{equation}%
The effective Hamiltonian $\mathcal{H}_{\text{BT}}\ $depicts a period$-3$ or
trimerized chain system, which can be further brought to a diagonal form%
\begin{equation}
\mathcal{H}_{\text{BT}}=\sum_{k}\Lambda _{k}\left\vert k\right\rangle
\left\langle k\right\vert ,
\end{equation}%
where $k=2\pi n/N$, $n\in \left[ 1,N\right] $ is the momentum and $%
\left\vert k\right\rangle $\ is the eigenstate of $\mathcal{H}_{\text{BT}}$.
Here we have used the Fourier transformation%
\begin{equation}
\underline{\left\vert \lambda ,k\right\rangle }=\frac{1}{\sqrt{N}}%
\sum_{l=1}^{N}e^{-ikl}\underline{\left\vert 3l-\lambda \right\rangle },
\end{equation}%
to construct the secular equation%
\begin{equation}
\left(
\begin{array}{ccc}
0 & -\sqrt{2}t & -2te^{-ik} \\
-\sqrt{2}t & 0 & -\sqrt{2}t \\
-2te^{ik} & -\sqrt{2}t & 0%
\end{array}%
\right) \left(
\begin{array}{c}
\gamma _{1,k} \\
\gamma _{2,k} \\
\gamma _{3,k}%
\end{array}%
\right) =\Lambda _{k}\left(
\begin{array}{c}
\gamma _{1,k} \\
\gamma _{2,k} \\
\gamma _{3,k}%
\end{array}%
\right)
\end{equation}%
for the eigenstate $\left\vert k\right\rangle =\sum_{\lambda =1}^{3}\gamma
_{\lambda ,k}\underline{\left\vert \lambda ,k\right\rangle }$. The spectrum $%
\Lambda _{k}$\ is solution of $\Lambda _{k}^{3}-8t^{2}\Lambda
_{k}+8t^{3}\cos k=0$, which possesses gaps at $\pm \pi /3$ and $\pm 2\pi /3$
due to the trimerization. Nevertheless, in the case of the weak
trimerization, the eigenstates around $\pm \pi /2$\ are hardly influenced by
the trimerization. Then the Gaussian wave packet of the form (\ref{Gaussian
WP}) with $k_{0}=\pm \pi /2$ should be still fast and robust against the
spreading. The corresponding optimal three-particle wavepacket is denoted as
$\left\vert \chi (\pm ,N_{c})\right\rangle $.

%%%%%%%%%%%%%%%%%%%%%%%%%%%%%%%%%%%%%%%%%%%%%%%%%%%%%%%%%%%%%%%%%%%%%%%%
\begin{figure}[tbp]
\includegraphics[ bb=89 349 505 720, width=6.0 cm, clip]{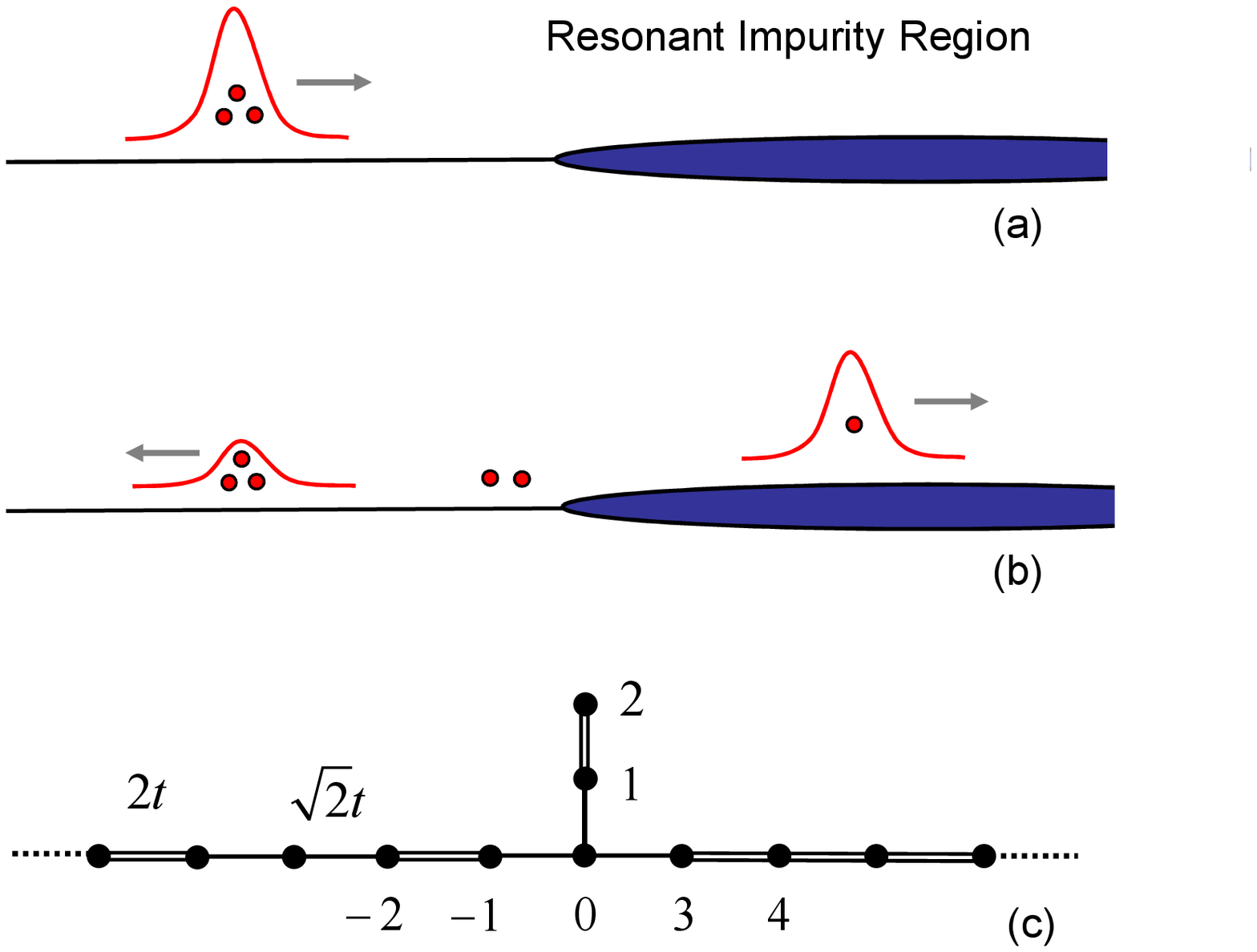}
\caption{(Color online) Schematic illustration of the coherent separation
process of a BT in the proposed optimal system. It consists of two
semi-infinite uniform chains with different chemical potentials, resonant
impurity. (a) An incident BT wavepacket comes from the left and scatters off
the barrier. (b) The reflected waves contain the BT and the single particle
wavepackets with different speeds. A single particle wavepacket moves along
the resonant impurity region. (c) Schematic illustration of the equivalent
effective Hamiltonian governs the evolution. For an optimal incident
wavepacket, the success probability of the coherent separation approaches to
unity. }
\label{Trimer}
\end{figure}

%%%%%%%%%%%%%%%%%%%%%%%%%%%%%%%%%%%%%%%%%%%%%%%%%%%%%%%%%%%%%%%%%%%%%%%%

The scheme to perform the coherent separation of a BT is based on the system
depicted by the Hamiltonian

\begin{eqnarray}
H_{\text{BTS}} &=&-t\left( \overset{0}{\sum_{i=-\infty }}+2\overset{\infty }{%
\sum_{i=2}}\right) a_{i}^{\dag }a_{i+1}-\sqrt{2}ta_{1}^{\dag }a_{2}+\text{%
H.c.}  \notag \\
&&+V\overset{\infty }{\sum_{i=-\infty }}n_{i}n_{i+1}+V\overset{\infty }{%
\sum_{i=2}}n_{i}.
\end{eqnarray}%
To investigate the scattering process of an input BT wave from the left, one
can establish an effective Hamiltonian in the quasi-invariant subspace
spanned by the following basis $\{\underline{\left\vert l\right\rangle }%
,l\in \left( -\infty ,\infty \right) \}$ defined as

\begin{equation}
\underline{\left\vert l\right\rangle }\equiv \left\{
\begin{array}{c}
\text{Eqs. (\ref{Tri_basis}), }\left( l\leqslant 2\right) \\
a_{-1}^{\dag }a_{0}^{\dag }a_{l-1}^{\dag }/\sqrt{2}\left\vert \text{vac}%
\right\rangle \text{, }\left( l>2\right)%
\end{array}%
\right. .
\end{equation}%
The corresponding equivalent effective Hamiltonian can be written as%
\begin{eqnarray}
\mathcal{H}_{\text{BTS}} &=&-\sqrt{2}t\left( \sum_{l=-\infty }^{2}+\sqrt{2}%
\overset{+\infty }{\sum_{l=4}}\right) \underline{\left\vert l-1\right\rangle
}\underline{\left\langle l\right\vert }-\sqrt{2}t\underline{\left\vert
0\right\rangle }\underline{\left\langle 3\right\vert }  \notag \\
&&-\left( 2-\sqrt{2}\right) t\overset{0}{\sum_{j=-\infty }}\underline{%
\left\vert 3j+1\right\rangle }\underline{\left\langle 3j+2\right\vert }+%
\text{H.c.}
\end{eqnarray}%
which is illustrated in Fig. \ref{Trimer} (c). The equivalent system is two
connected semi-infinite chains with a side coupled two-site segment at the
joint of them.\ In the optimal case, the perfect separation (or combination)
process in the real space can be expressed as

\begin{equation}
\begin{split}
& \left\vert \chi (+,-\infty )\right\rangle \left\vert \text{vac}%
\right\rangle \Longrightarrow \\
& r\left\vert \chi (-,-\infty )\right\rangle \left\vert \text{vac}%
\right\rangle +t\left\vert \varphi (+,+\infty )\right\rangle a_{-1}^{\dag
}a_{0}^{\dag }\left\vert \text{vac}\right\rangle ,
\end{split}%
\end{equation}%
and the corresponding inverse process as

\begin{equation}
\begin{split}
& \left\vert \varphi (-,+\infty )\right\rangle a_{-1}^{\dag }a_{0}^{\dag
}\left\vert \text{vac}\right\rangle \Longrightarrow \\
& r\left\vert \varphi (+,+\infty )\right\rangle a_{-1}^{\dag }a_{0}^{\dag
}\left\vert \text{vac}\right\rangle +t\left\vert \chi (-,-\infty
)\right\rangle \left\vert \text{vac}\right\rangle .
\end{split}%
\end{equation}%
This separation process is a little different from that of the BP. An
incident three-particle wavepacket leaves two NN pair at the impurity, while
a single-particle remains going forward. Numerical simulation is performed
to evaluate the success probability of the previous scattering process. For
an optimal Gaussian wavepacket with $\alpha =0.01$ the separation success
probability is $0.97$ approximately.

\section{Conclusion}

In summary, we studied the BP and BT states in the extended Hubbard model.
We found that, the resonant NN interaction can induce the BP and triple
states. They are distinct from the known ones in the previous works, as
their bandwidths are comparable to that of a single particle. In other
words, the bandwidths of the bound clusters can be drastically widened by
the NN interaction.\ We proposed relatively simple\ schemes to realize the
coherent separation of these bound clusters. The exact result showed that
the success probability of the coherent separation is unity for a BP while
approached to unity for a BT in the optimal systems. We also studied the
singlet BP in the extended Fermi-Hubbard system. We showed that the
corresponding coherent separation process can be utilized to create
long-distance entanglement without the need of temporal control and
measurement process. Finally, we believe that our study will shed more light
on the future research for multi-particle bound state.

We acknowledge the support of the CNSF (Grant Nos. 10874091 and
2006CB921205).

\end{document}